\documentclass[12pt]{iopart}
\usepackage{graphicx}

\begin{document}

\title[Argon metastable dynamics in a filamentary jet micro-discharge]{Argon metastable dynamics in a filamentary jet micro-discharge at atmospheric pressure}

\author{B Niermann, R Reuter, T Kuschel, J Benedikt, M B\"oke and J Winter}

\address{Ruhr-Universit\"at Bochum, Institute for Experimental Physics II, Universit\"atsstra\ss e 150, 44780 Bochum, Germany}

\ead{benedikt.niermann@rub.de}

\begin{abstract}

Space and time resolved concentrations of Ar ($^3\mbox{P}_2$) metastable atoms at the exit of an atmospheric pressure radio-frequency micro-plasma jet were measured using tunable diode laser absorption spectroscopy. The discharge features a coaxial geometry with a hollow capillary as an inner electrode and a ceramic tube with metal ring as outer electrode. Absorption profiles of metastable atoms as well as optical emission measurements reveal the dynamics and the filamentary structure of the discharge. The average spatial distribution of Ar metastables is characterized with and without a target in front of the jet, showing that the target potential and therewith the electric field distribution substantially changes the filaments' expansion. Together with the detailed analysis of the ignition phase and the discharge's behavior under pulsed operation, the results give an insight into the excitation and de-excitation mechanisms.

\end{abstract}

\pacs{52.30.-q, 52.70.-m, 52.35.-g, 52.40.Hf, 52.50.-b, 52.50.Dg}
\submitto{Plasma Sources Science and Technology}

\maketitle

\section{Introduction}

In recent years the research on atmospheric pressure micro-plasmas became a strong focus in plasma science, mostly due to their high potential for new plasma applications without the need for expensive vacuum equipment. Among the large variety of micro-plasma sources, that make use of DC, pulsed DC and AC voltages ranging from mains frequency to RF, a large number of configurations has been described for micro-discharges operated at atmospheric pressure. Examples are micro-hollow cathode discharges, capillary plasma electrode discharges, dielectric barrier discharges, and micro-plasma jets \cite{becker, Iza}.

Understanding the energy transfer processes in micro-plasmas is
one of the key requirements for developing new applications and a reliable
process control. In this context metastable species play a
decisive role. Due to their long lifetime metastables collide
more frequently with other particles. The metastable density
in micro-discharges is several orders of magnitude lower than
the density of the ground-state atoms. However, compared to
most other species the density is significant and the electron
collision excitation cross sections of some argon levels out
of the metastable states exhibit values which are several orders
of magnitude larger and have much lower thresholds than those
for the ground state \cite{Katsch1996, Flohr1993}.  Among the Ar metastable
states the Ar ($^3\mbox{P}_2$) level plays a decisive role, since
argon is used as a feed gas in many micro-discharges.

Reliable techniques are needed for the systematic investigation
of plasma characteristics and dynamics. This is essential for
the optimization of plasma sources and process control.
Application of conventional diagnostics to micro-plasmas, especially invasive
ones, is often impossible regarding the small dimensions,
high operating pressures, and high power densities. In this
context absorption spectroscopy is a widely used technique
to measure e.g. absolute concentrations of particles in
plasma discharges, since it is non-invasive, highly sensitive
and provides a sufficient spatial resolution \cite{Niemax2005,
Tachibana2005}.

Among the variety of micro-plasma sources, RF driven atmospheric pressure plasma jets provide an effective platform for various plasma chemical and surface reactive processes \cite{selwyn, schulz, benedikt, schaefer2}. Here we present the results for a micro-plasma jet with a hollow capillary as an inner electrode and a ceramic tube with metal ring as outer electrode. This design is already used for coating and surface treatment applications \cite{raballand, reuter}. While this jet design shows large potential in these domains, there is still a lack of understanding the fundamental discharge dynamics. We applied tunable diode laser absorption spectroscopy (TDLAS) to record the spectral line profile of the lowest argon metastable state ($^3\mbox{P}_2$) from the 1s$_5\rightarrow$2p$_9$ transition, deducing absolute densities for various discharge conditions. In addition to TDLAS we used optical emission spectroscopy with high temporal resolution to observe the discharge's filamentary structure.

\section{The atmospheric pressure micro-plasma jet}

\begin{figure}
    \centering
	\includegraphics[width=\textwidth]{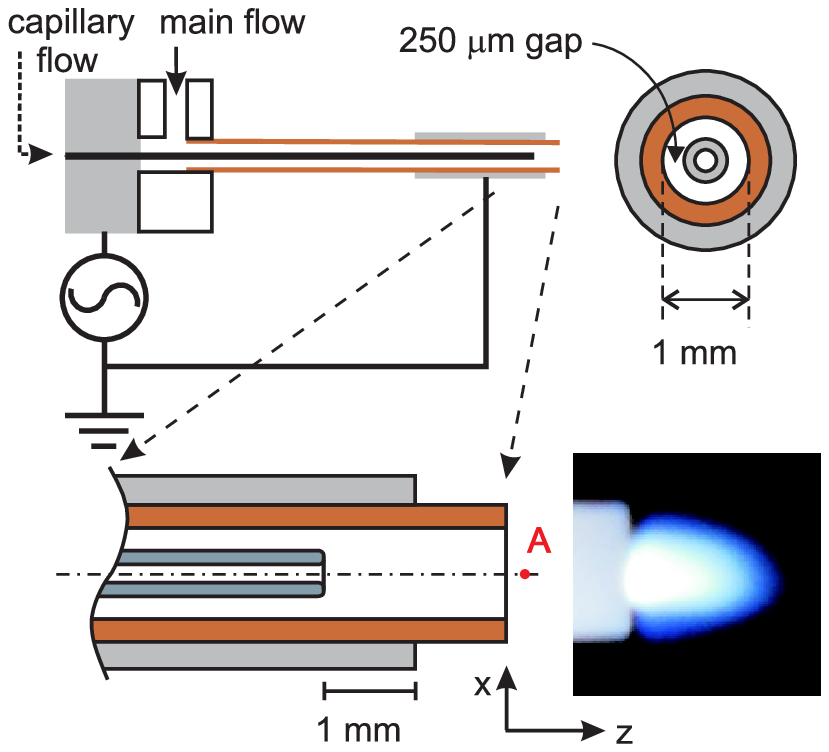}
	\caption{Scheme of the micro-plasma jet discharge with the photograph of the plasma effluent in the Ar/He atmosphere. The position A indicates the position of the laser beam for measurements presented in figures \ref{figure3} and \ref{figure4}.}
	\label{figure1}
\end{figure}

Figure \ref{figure1} shows a scheme of the microplasma jet. The detailed description of the setup has been published previously \cite{yanguas, benedikt2, benedikt3}. A stainless steel capillary tube is inserted into a ceramic tube leaving an annular gap of 250\,$\mu$m between
the tubes. The capillary ends 2\,mm prior to the end
of the ceramic tube. Around the ceramic tube and 1\,mm
apart from its end, an aluminum tube serves as counter
electrode. A 13.56 MHz radio frequency power supply can
be attached to the capillary
or to the outer electrode through a matching network. Plasma forming gas (He or Ar in this case) is introduced
into the annular space between the ceramic tube
and the capillary with a flow rate of 3000\,sccm (main flow). Additional a
flow of typically 160\,sccm (capillary flow) is guided through the capillary in order
to maintain similar gas velocities in the annular space between
electrodes and in the capillary. This is especially
important when some reactive gas is added to the capillary
flow. The plasma can be ignited easily in He by
applying root mean square (RMS) voltage of about 100 V.
The ignition in Ar is not possible without an external high
voltage pulse and therefore an Ar-plasma operation is realized
through ignition in He and switching gas flows to
Ar. Nevertheless it is possible to operate the plasma pulsed, as long as the power off time is less than 8\,$\mu$s.

The whole jet setup is located in an airtight stainless steel vessel, whose atmosphere was filled with helium and argon mixture at a ratio of 5:3 after pumping it to a base pressure of 1\,Pa. This gas mixture has been chosen to correspond to experimental conditions used in another experiment for surface treatment \cite{reuter}.

To observe the ignition phase of the jet, the plasma was operated in pulsed mode, with a frequency of 1\,kHz and a duty cycle of 99.4\,\%. The short power off time is sufficient due to short metastable lifetimes in the order of some 100\,ns. A steady state condition of the discharge is reached about 150\,$\mu$s after the ignition. The plasma extends up to 2\,mm from the exit nozzle and looks homogeneous to the eye, but has strong filamentary structure, similar to the results described by Sch\"afer et al. \cite{Schaefer}, as will be shown in this article.

\section{Spectroscopic setup}

\begin{figure}
    \centering
	\includegraphics[width=\textwidth]{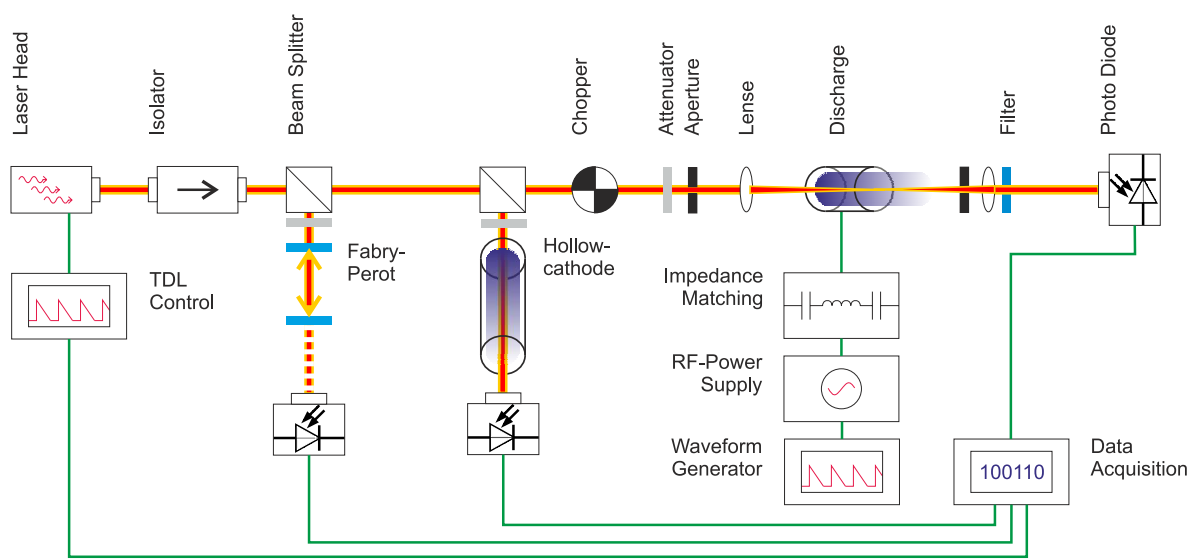}
	\caption{TDLAS setup for absolute argon metastable density measurements.}
	\label{figure2}
\end{figure}

Figure \ref{figure2} shows a sketch of the spectroscopic setup. The laser beam from the diode laser is guided through an optical isolator, to protect the diode from back-reflections, and passes two beam splitters to create three equivalent branches.

The first branch leads to a low pressure reference cell. In our case a hollow cathode discharges filled with argon. The reference cell provides longer absorption length and since it is running at low pressures around a few tens millibar, the lines are narrow and dominated by the Doppler width of the line profile. A second branch is guided to a confocal Fabry-Perot interferometer with 1\,GHz free spectral range, necessary for the frequency calibration of the spectrum.

The part of the beam transmitted through the first beam splitters is attenuated by neutral density filters with an optical density in the order of 3, and focused into the discharge with a beam power of less than 2\,$\mu$W at 100\,$\mu$m spot size, to avoid any saturation effects. After passing the discharge the beam was guided through a set of apertures and filters to suppress the emission from the plasma by reducing the collection angle and blocking wavelengths different than the observed transitions. The transmitted beam intensity was measured by very fast low noise photodiodes for highly time resolved measurements. The laser beam was focused into the discharge with a focal lengths of 4\,cm. Assuming ideal lenses and taking into account the aperture size of 2\,mm and the used wavelength of 811\,nm, the spot size of the beam in the focal point can be calculated to be about 30\,$\mu$m. In the real system measurements show that the spot size is around 100\,$\mu$m. The time resolution of the system is about 40\,ns. For an effective measurement of the absorption signal across the jet axes, the discharge casing was mounted on a small movable stage featuring three electronically controlled stepping motors to adjust the plasma jet position in all spatial dimensions with high precision. This setup allows the positioning of the jet with an accuracy of about 30\,$\mu$m. All measurements presented are line-integrated in y-axis. The time averaged measurements presented in this article are accumulated over 500 pulse cycles.

Absolute metastable densities were derived from the transmittance $\frac{I}{I_0}$ and the Beer-Lambert law. Therefore four signals had to be acquired:
\begin{eqnarray*}
&L(\nu)\, -& \mbox{Plasma and Laser on},\\
&L_0(\nu)\, -& \mbox{Plasma off, Laser on},\\
&P(\nu)\, -& \mbox{Plasma on, Laser off},\\
&B(\nu)\, -& \mbox{Plasma and Laser off (background)},
\end{eqnarray*}
to calculate the transmittance spectra and correlate them with the plasma properties by
\begin{eqnarray}
\frac{I(\nu)}{I_0(\nu)}=\frac{L(\nu)-P(\nu)}{L_0(\nu)-B(\nu)}=e^{-k(\nu)\cdot l},
\end{eqnarray}
where $I(\nu)$ and $I_0(\nu)$ are the intensities of transmitted radiation with and without the presence of absorbing species, $k(\nu)$ is the absorption coefficient and $l$ the path length through the absorbing medium \cite{Sadeghi}. The absorption coefficient is connected to the population density of metastable atoms by
\begin{eqnarray}
k(\nu)=\frac{1}{4\pi\epsilon_0}\cdot\frac{\pi(e^2)}{c m_e }\cdot\ f_{ik}\cdot N_i \cdot F(\nu),
\end{eqnarray}
where $f_{ik}$ is the oscillator strength of the line, $N_i$ the density of the lower level, and $F(\nu)$ a normalized function ($\int_{0}^\infty F(\nu)\cdot d\nu=1$) representing the absorption line-shape. All other terms have their usual definitions.
The absolute metastable density can then be given by
\begin{eqnarray}
\int_0^\infty\ln\left ( \frac{I_0(\nu)}{I(\nu)}\right )d\nu=S=\frac{e^2f_{ik}l}{4\epsilon_0 m_e c}\cdot  N_i,
\end{eqnarray}
where $S$ is the area under the absorption curve that provides the line of sight-averaged density of the absorbing species. To derive absolute values for the metastable density, the spectral profile of the lines has to be analyzed and the pressure broadening as well as the peak intensities have to be measured. The area of the absorption line profile is given by
\begin{eqnarray}
S&=&\int_0^\infty\ln\left ( \frac{I_0(\nu)}{I(\nu)}\right )d\nu=-\int_0^\infty\ln (1-A(\nu))d\nu,
\end{eqnarray}
where $A(\nu)=\frac{I_0(\nu)-I(\nu)}{I_0(\nu)}$ is the absorption rate at frequency $\nu$ and $\nu_c$ is the central frequency of the absorption line. The absorption line-shape can be described by a Voigt function, i.e., a convolution of lorentzian and gaussian functions. The width of the gaussian component $\Delta\nu_D$, caused by Doppler broadening is described by
\begin{eqnarray}
\Delta\nu_D=\frac{2}{\lambda_0}\cdot\sqrt{2ln(2)\frac{k_b T}{M}},\label{dopplerbroadening}
\end{eqnarray}
where $T$ is the temperature of the absorbing species (here assumed equal to the gas temperature) and M the mass of the absorbing species \cite{Demtroeder}. A more detailed insight into the density calculations is given in \cite{niermann}.

Additionally the light emission profile of the discharge effluent is recorded with an ICCD camera (Andor, iStar DH734-18F-03) in the visible spectral range. The experiment and the camera are synchronized with the internal gate monitor of the ICCD camera.

\section{Results and discussion}

\subsection{Density approximations and experimental uncertainties}

\begin{figure}
    \centering
	\includegraphics[width=\textwidth]{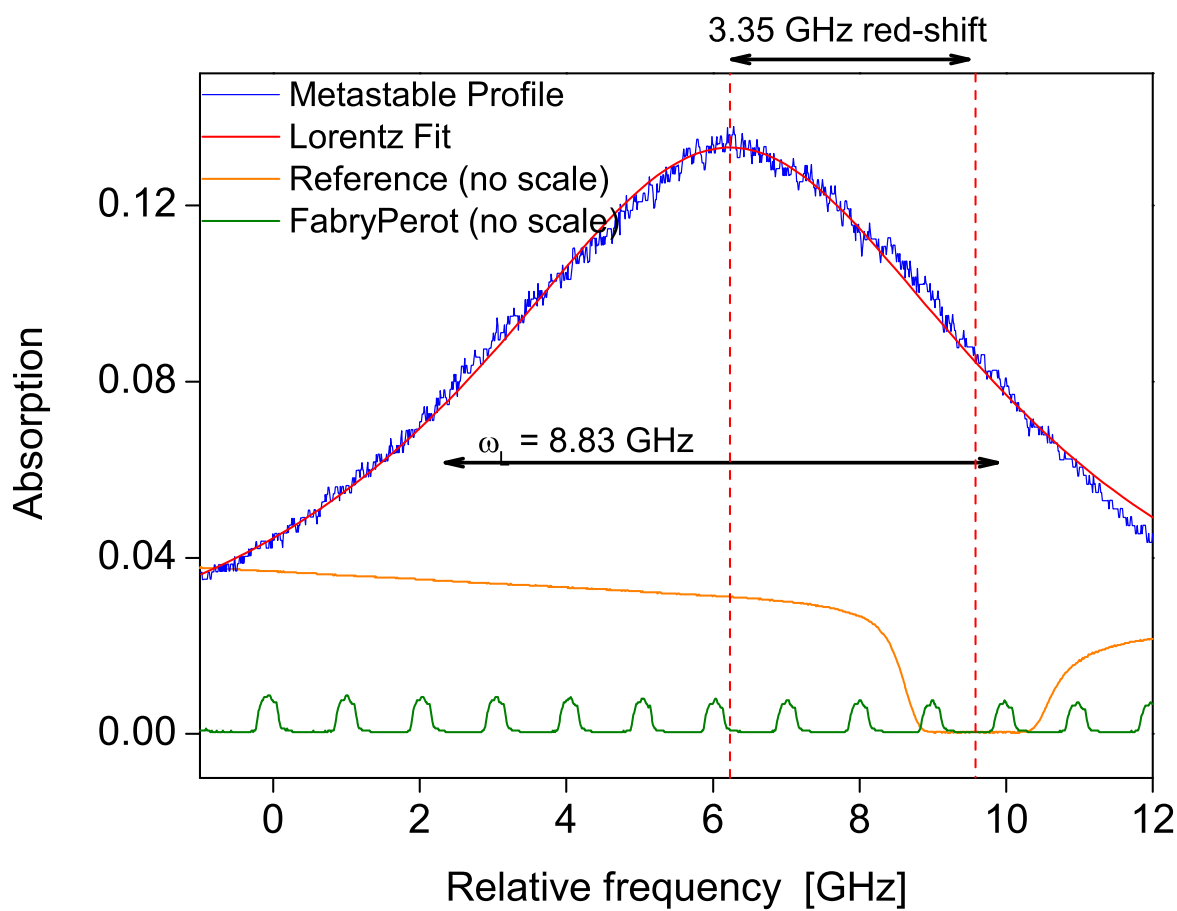}
	\caption{Spectral absorption profile of the argon metastable 1s$_5\rightarrow$2p$_9$ transition, showing strong lorentzian broadening and significant red shift of the line. Shown are also the signals from the low pressure hollow cathode lamp (reference) and the Fabry-Perot interferometer.}
	\label{figure3}
\end{figure}

Figure \ref{figure3} shows the time averaged spectral profile of the absorbing transition, measured about 200\,$\mu$m behind the nozzle (position A in figure \ref{figure1}) under steady state conditions about 150\,$\mu$s after the plasma ignition. The line shows dominating Lorentzian broadening with a half width of about 8.8\,GHz. The line is significantly shifted compared to its position under low pressure and low electron density conditions. The shift is about 3.3\,GHz to longer wavelength, which is almost 2\,GHz more than in $\alpha$-mode glow discharges at atmospheric pressure \cite{niermann}. Due to high electron densities in the filaments, the shift must probably be attributed to Stark effects. Since the discharge is strongly fluctuating, both in time and space, the spectral profile is expected not to be constant. The time averaged profile shape was used to calculate the absolute metastable densities, because accounting for the real absorption profile with a temporal and spatial resolution of nanoseconds and micrometers was not in the realm of possibility. The wavelength where the absorption profile has its maximum is used in the following for the space and time resolved measurements.

Since the filaments are not confined to a certain spatial position, a time resolved measurement of the spectral profile is not accomplishable, thus estimations of the electron densities in the filament by analyzing the Stark shift of the spectrum is not achievable. The strongly dominating Lorentzian part of the profile, covering the Doppler width, also prohibited the calculation of gas temperatures with reasonable accuracy.\\

Although the absorption can be measured very accurately, the fluctuating and inhomogeneous character of the discharge affects the precise calculation of absolute densities in this case, and makes an evaluation of the errors difficult. The spectral profile shown in figure \ref{figure3} is the result of an averaged measurement 200\,$\mu$m behind the nozzle. But due to the filamentary type of the discharge, the spectral profile is expected to change significantly between inside and outside the filaments. Also the gas temperature, that is changing along the plasma effluent, determines the broadening and the shape of the profile. Small changes in the width of the line can have significant effects on the calculated metastable densities. A second strong impact is given by the absorption length, that is different for any position in the effluent. Furthermore are all measurements line of sight integrated. Applying an Abel transformation to the measurements was not feasible taking into account that the resulting values are still strongly affected by the other effects limiting the accuracy of the calculated densities. Furthermore the Abel transformed profiles would have limited meaning, since they do not take into account the filamentary nature of the discharge. A third factor, affecting all averaged results shown in this section, lies in the order of data analysis. Since the signal is fluctuating strongly over the time, and relation between absorption and metastable density is not linear, it would be necessary to measure first a large amount of time resolved spectra and apply afterwards the averaging procedure. However, due to the complexity in post-processing the data, this order was not feasible, and the absorption signal was first averaged and then converted into absolute densities.

Taking these circumstances into account the graphs shown in this article provide mostly data about the measured absorption instead of absolute densities. If density information are given, they have been calculated with an absorption length that was taken as the half width of the vertical plasma beam profile.

\subsection{Non-averaged signals}
\label{section:nonaveraged}

\begin{figure}[H]
    \centering
    \includegraphics[width=\textwidth]{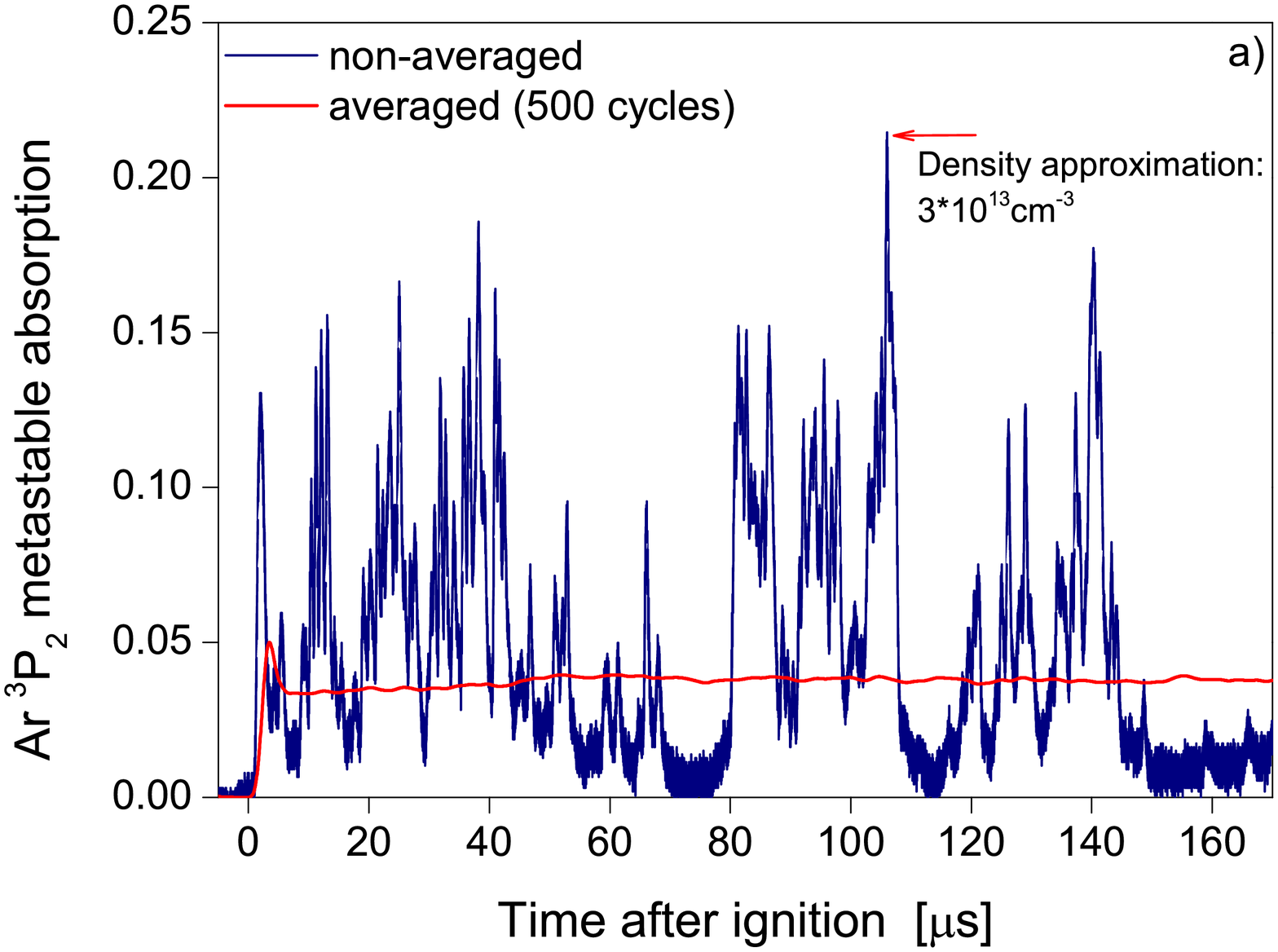}
    \includegraphics[width=\textwidth]{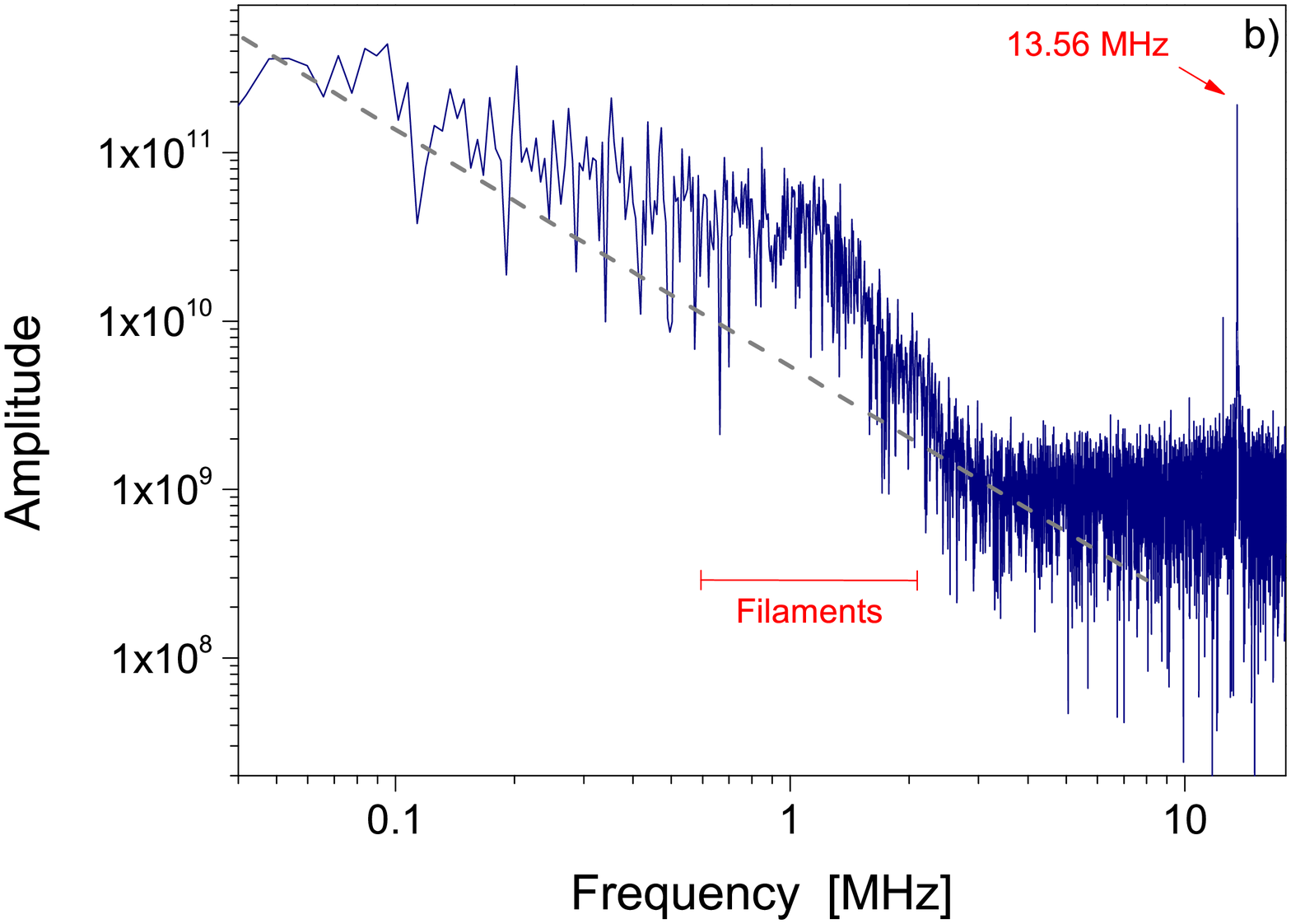}
     \caption{\textbf{a):} Non-averaged (blue) and averaged (red) time resolved measurement of argon metastables in the first 160\,$\mu$s after the ignition of the jet, taken on the jet axis, about 200\,$\mu$m in front of the nozzle. \textbf{b):} Fourier analysis of the temporal density evolution, showing the metastable events in a strongly broadened peak around 1\,GHz.}
     \label{figure4}
\end{figure}

Figure \ref{figure4} a) shows a non-averaged time resolved measurement of argon metastables in the first 160\,$\mu$s after the ignition of the jet. It was taken on the jet axis, about 200\,$\mu$m in front of the nozzle (position A in figure \ref{figure1}). The measurements reveal the filamentary nature of the discharge by showing a strong peak in the spectrum whenever a filament crosses the laser beam. Outside the filaments the absorption is more than one order of magnitude lower. Taking this into account, we can conclude that only the high density in the filaments contributes to the spectral absorption profile shown before. Together with an estimation of the absorption length in one filament, that is given at the end of this section, the density inside a filament can be approximated to be at least in the order of 10$^{13}$\,cm$^{-3}$. This is in agreement with the results obtained by Sch\"afer et al. \cite{Schaefer}.

Figure \ref{figure4} b) shows a Fourier analysis of the temporal density evolution. Beside the random noise, that would appear as a brownian 1/f$^2$ spectrum, the frequency spectrum shows two anomalies. One is the peak at 13.56\,MHz caused by the RF frequency used to power the jet discharge. The other is a broad peak shaped structure, distorting the spectrum at frequencies around 1\,MHz. This broad peak represents the characteristic frequency range of the filaments crossing the laser beam. It is determined by a convolution of several factors given by the spot-size of the laser focus, rise and decay times of the metastable atoms and the lifetime of the filament itself. The strong broadening of the frequency peak is caused by a variety of factors: Fluctuations in the gas mixture, the gas temperature, as well as quenching effects by high electron and metastable densities inside the filaments strongly influencing the species lifetime. The pile up of subsequent filaments passing the point of measurement leads to a smearing of the frequencies to lower values.

Three-body collisions are the dominating effect for limiting the lifetime at atmospheric pressure, but for argon species the rate coefficient is known to have a strong dependance on the gas temperature as well. The average gas temperature at the exit of the nozzle is well below 350\,K, but is expected to show strong temporal and spatial fluctuations due to the filamentary nature of the discharge. Furthermore the spatial temperature profile in the effluent of the discharge is expected to have strong gradients due to the dilution of the feed gas stream with gas from the ambient atmosphere, which is in large parts helium. Inside the filaments, both electron and metastable densities become high enough to show significant impact on metastable quenching by superelastic collisions and pooling reactions. Additionally, nitrogen, oxygen as well as water molecules are desorbing from the gas supply system or the chamber walls and intermix with the working gas, so that the ambient atmosphere is estimated to contain at least some dozens of ppm of air \cite{niermann2}. All these effects make a theoretical estimation of the different factors that limit the metastable lifetime unfeasible.

\begin{figure}[H]
    \centering
	\includegraphics[width=\textwidth]{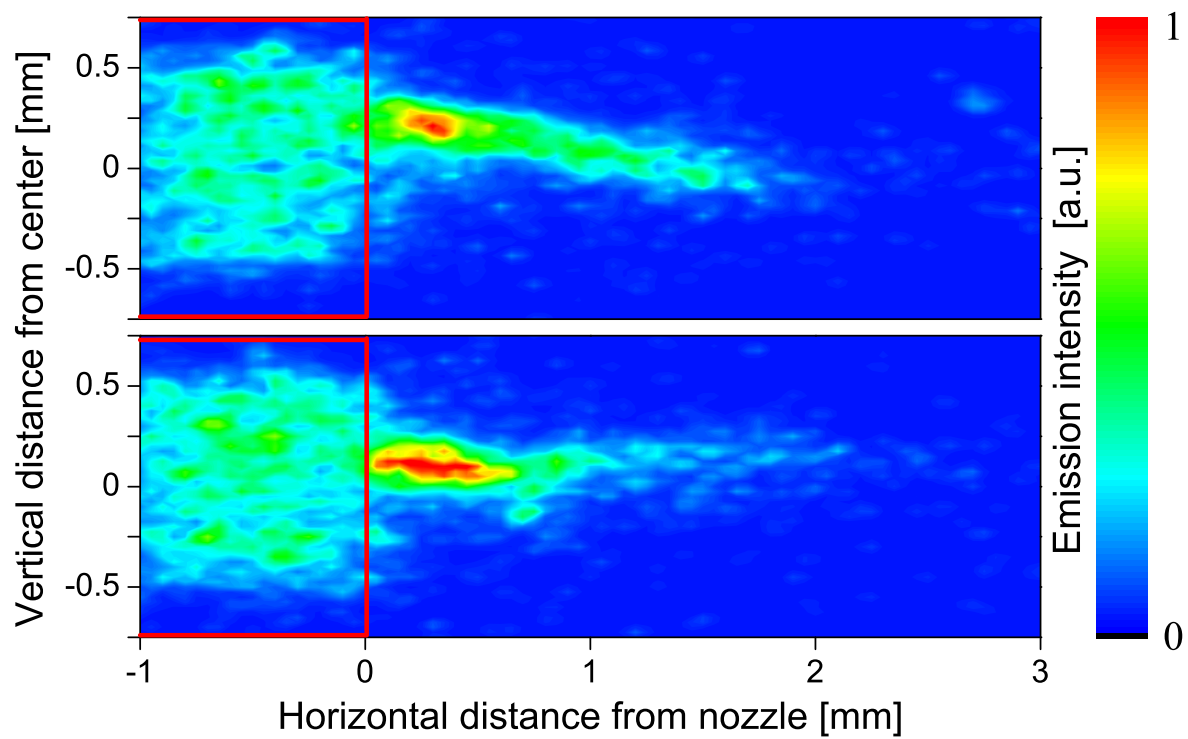}
	\caption{Emission profiles of the effluent region and the capillary in the visible range, measured with an exposure time of 1\,$\mu$s. The red boxed area indicates the position of the ceramic capillary. The light emissions from the surface of the ceramic tube is plasma emission originating from the filaments inside the tube. Radiation gets scattered in all directions when passing the ceramic.}
	\label{figure5}
\end{figure}

\begin{figure}
    \centering
	\includegraphics[width=\textwidth]{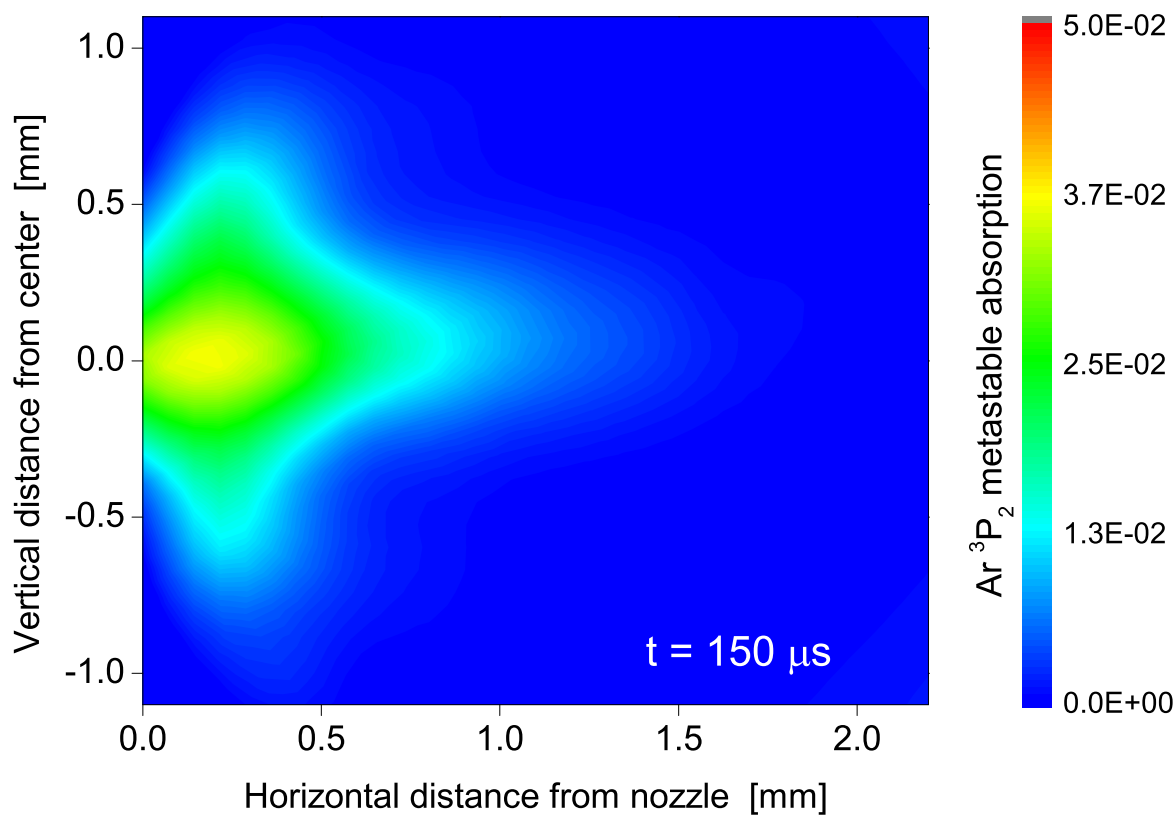}
	\caption{Averaged steady state absorption map of the argon metastable distribution, measured 150\,$\mu$s after the ignition of the jet behind the nozzle.}
	\label{figure6}
\end{figure}

To confirm the assumption of a filamentary discharge, fast optical emission measurements were used to record the emission profiles in the visible range with a high temporal resolution in the order of one nanosecond. As indicated by the metastable measurements the movement of the filaments is slow, in the order of 100\,ms$^{-1}$. Figure \ref{figure5} shows two typical emission maps of the effluent and the capillary in the visual range. The maps are time integrated over 1\,$\mu$s, still providing sufficient spatial confinement due to the slow movement of the filaments. The upper diagram shows the typical structure of a filament. It originates from the inner powered metal electrode and extends to about 2\,mm distance from the nozzle. The course of the filament is typically slightly curved, but it constantly tends to head back to the central axis. This may be due to the electric field distribution in the effluent or the fact that in the central channel the working gas is mostly unaffected by the ambient gas environment and the propagation for the filament is easiest here. Comparing this bended structure with the profile of the metastables shown in the next section, the course of the filaments explains the star-shaped excitation profile in the effluent, that is shown in figure \ref{figure6}. The diameter of the filaments is about 125\,$\mu$m, which is in good agreement with the values modeled by Sch\"afer et al. \cite{Schaefer}.

\subsection{Spatial profiles in the continuous plasma}

Figure \ref{figure6} shows the averaged steady state metastable distribution behind the nozzle that is reached in the steady-state condition about 150\,$\mu$s after the ignition of the jet. The measurements have been obtained by averaging the signal over 500 pulse cycle. The map shows significant densities far outside the capillary, although the gas velocity on the exit of the nozzle is slightly below 100\,ms$^{-1}$ and lifetimes are in the order of a few hundred nanoseconds. This limits the decay length of this species to some tens of micrometers without further excitation. Taking this into account, the density distribution suggests significant electron densities and temperatures in more than 1.5\,mm distance from the nozzle, leading to high production rates of metastable atoms. This observation supports our previous suggestion, based on the phase resolved emission spectroscopy data, that the discharge has a streamer-like character and can be filamentary \cite{benedikt}.

The plasma burns on the average symmetrically and extends in vertical direction over a distance of almost 2\,mm. The star-shaped expansion of the metastable profile is conditioned by the propagation course of single filaments, that was discussed in section \ref{section:nonaveraged}.

\subsection{Spatial profiles during the ignition phase}

\begin{figure}
    \centering
	\includegraphics[width=\textwidth]{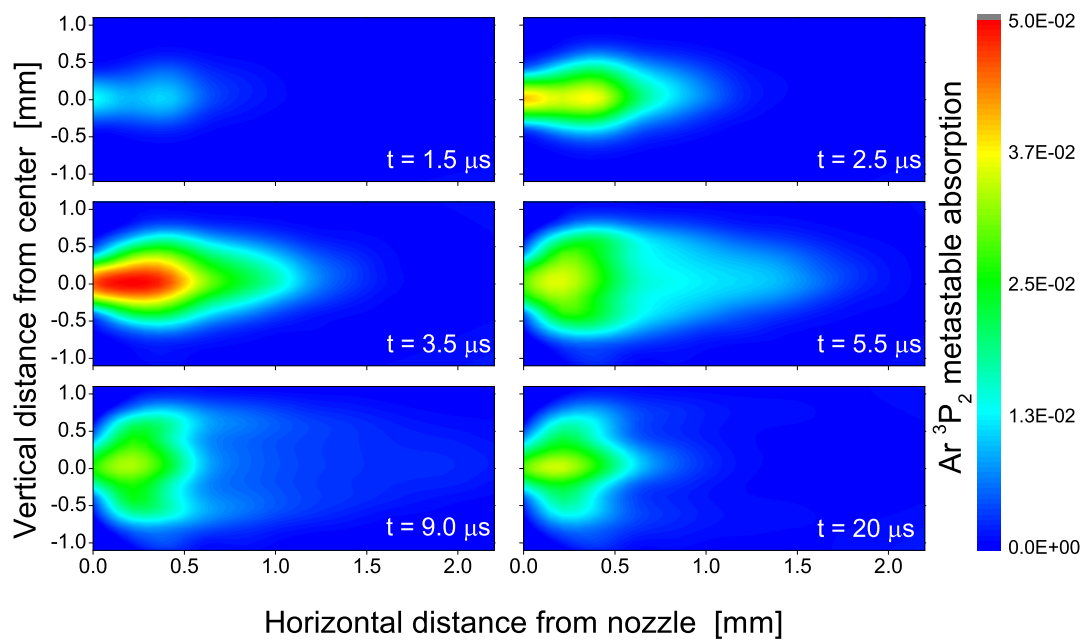}
	\caption{Absorption maps of argon metastable distributions for selected points in time, representing the ignition phase of the jet in the first 20\,$\mu$s.}
	\label{figure7}
\end{figure}

Since the absorption was measured with a time resolution of about 40\,ns the ignition phase of the jet could be observed in detail. Figure \ref{figure7} shows an area of 2.2 x 2.2\,mm$^2$ behind the nozzle at certain times after the ignition. The graphs show major steps in the temporal evolution of the discharge. In the first 2\,$\mu$s the discharge forms in a channel of roughly 1\,mm diameter width at the tip of the nozzle. After 3.5\,$\mu$s a maximum in the metastable absorption is reached, located about 200\,$\mu$m in front of the nozzle. In the following 7\,$\mu$s the discharge expands over about 2\,mm in vertical direction and produces significant metastable densities in up to 2\,mm horizontal distance. The discharge reaches a steady state distribution about 150\,$\mu$s after the breakdown.

These changes in the temporal evolution may be an interesting observation for applications like surface treatment processes, where the materials are in some distance to the nozzle. If in the first breakdown phase of the discharge a significantly higher averaged metastable density can be achieved far away from the nozzle, a high frequency pulsing of the discharge may lead to improved results.

\subsection{Excitation waves}
\label{ExcitationWave}

\begin{figure}
    \centering
    \includegraphics[width=\textwidth]{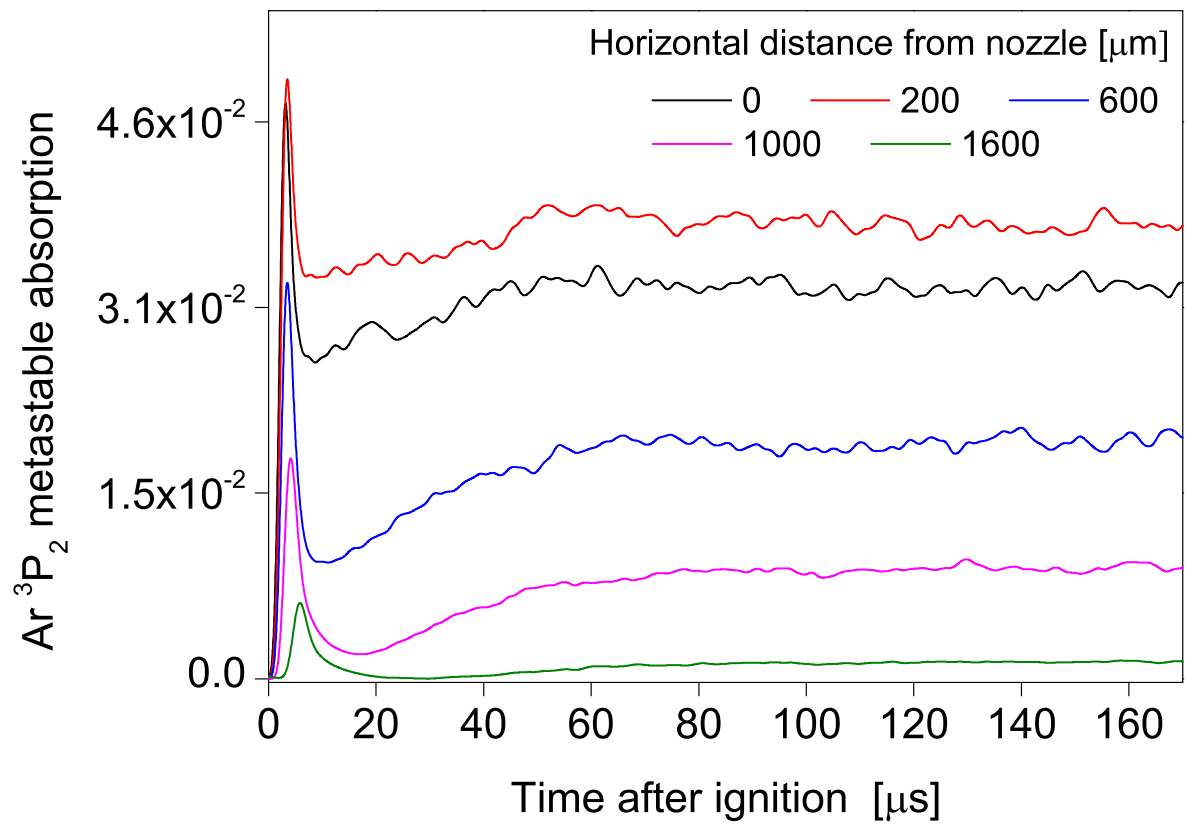}
   \caption{Averaged metastable absorption in the first 160\,$\mu$s after the ignition, measured on the central axis of the jet for various distances from the nozzle. }
     \label{figure8}
\end{figure}

While the temporal evolution of each cycle is highly chaotic, averaging the signal over about 500 cycles results in a stable signal. Figure \ref{figure8} shows the averaged metastable absorption in the first 160\,$\mu$s after the ignition, measured on the central axis of the jet at various distances from the nozzle. Although every new cycle shows an unpredictable sequence of filaments, there is a single commonality between each of them. In the first microseconds of the cycle a breakdown occurs close to the nozzle and travels outwards. As shown in figure \ref{figure4} a) the first filament in the filament train is not stronger than the subsequent ones, but since it has an almost fixed position in the time domain, the signal sums up in the averaging process and gives the impression of an enhanced metastable production just after the breakdown (see the averaged red absorption signal in figure \ref{figure4} a)). Furthermore the averaged signal shows a strong decrease after the first maximum, suggesting that the passing of a filament in a certain point of space seems to lower the probability for the passing of an additional one shortly after. Measurements of the current and voltage waveforms taken on the jet electrodes during the ignition phase indicate that the observations described above can not be explained by an overshoot of the power supply. Thus the averaged temporal metastable distribution gives valuable information about the statistical distribution of the filaments in space and time.

Measuring the temporal profile in dependance of the distance to the nozzle (fig. \ref{figure8}) opens the possibility to study the expansion of the discharge in the first microseconds after ignition. In this case the velocity was measured on the central axis of the jet, resulting in velocities of more than 1500\,ms$^{-1}$ just behind the nozzle, and less than 400\,ms$^{-1}$ in a distance of 1.6\,mm. The speed of sound in argon and helium is about 300 and 1000\,ms$^{-1}$ respectively and can therefore not be responsible for the observed velocity profiles. The expansion seems to follow an excitation wave as it was described by Kong et al. for a similar plasma jet discharge \cite{kong}.

\subsection{Metastable distribution in front of targets}

\begin{figure}
    \centering
	\includegraphics[width=\textwidth]{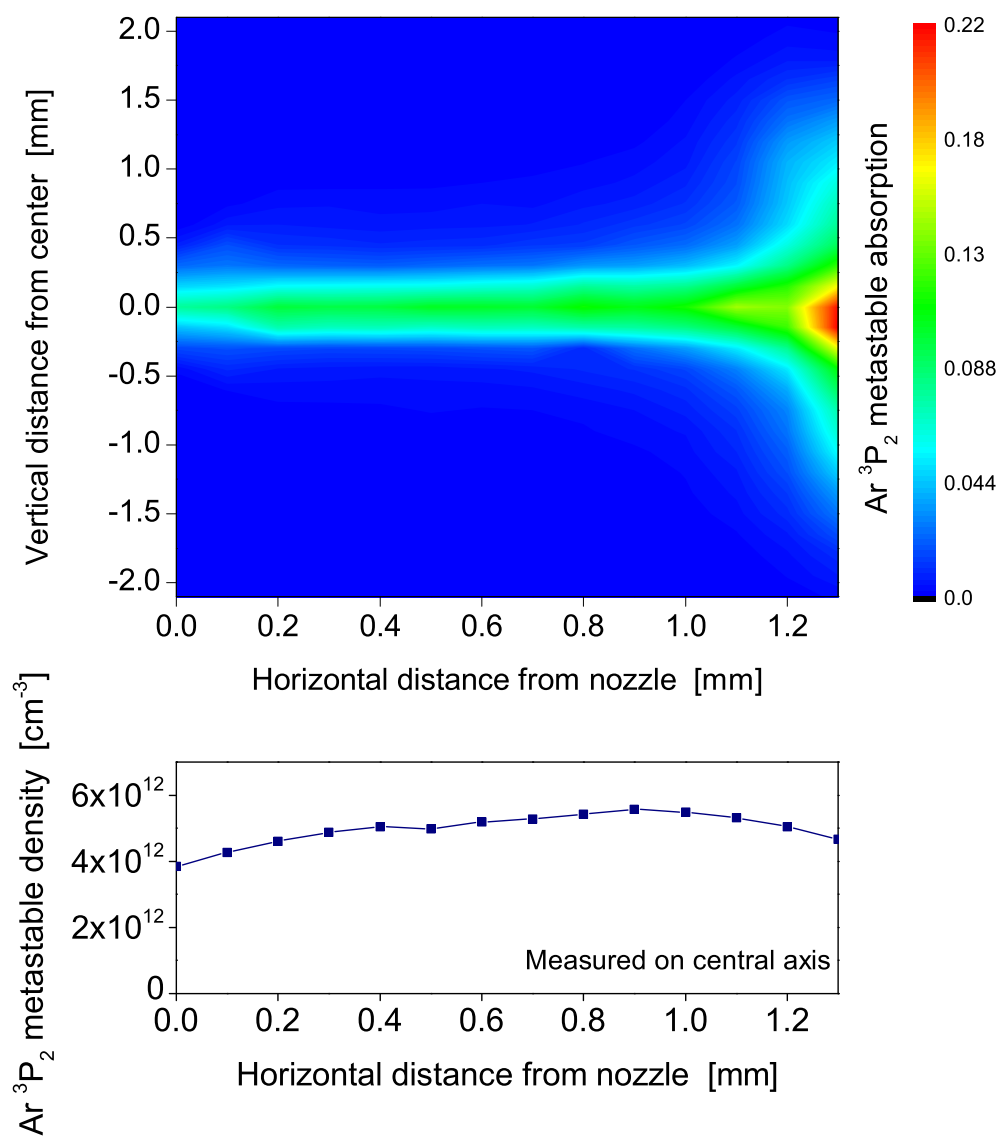}
	\caption{\textbf{Top:} Steady state absorption map of argon metastables between the nozzle and a grounded target in 1.3\,mm distance from the jet. \textbf{Bottom:} Steady state density distribution on the central axis between the nozzle and a grounded target in 1.3\,mm distance from the jet.}
	\label{figure9}
\end{figure}

One way to significantly change the excitation profile of metastables in the effluent is to introduce a target in a short distance from the nozzle. These measurements are in particular of interest to draw connections to surface treatment experiments that have been performed with this jet \cite{raballand}. The target is made from a silicon wafer that was mounted on a metal plate. A critical element in this context is the electrical potential of the metal plate directly behind the wafer. While an electrically floating plate leads to a density distribution that is almost identical to the situation without a target, grounding the metal parts significantly affected the characteristics of the discharge. Figure \ref{figure9} (top) shows the averaged steady state spatial distribution between the nozzle and a target in 1.3\,mm distance. The spatial profiles differ substantially to the measurements without target and floating target. The discharge shows a significantly more stable behavior and is not fully governed by the filamentary character observed in the free effluent. High metastable densities can be maintained over the whole plasma column between nozzle and target. The plasma column is more confined with a tighter vertical profile, probably caused by the electric field distribution that pulls the filaments on the target electrode. The star-shaped profile directly behind that nozzle is lost. Close to the targets surface the discharge spreads over a diameter of more than 3\,mm. Figure \ref{figure9} (bottom) gives absolute metastable densities on the central axis between the nozzle of the jet and the target. Densities have been calculated by taking the full width at half maximum of the measured absorption profile as absorption length from the spatial profile.

Under grounded target conditions the plasma column can be extended over a distance of about 3\,mm, still having averaged metastable densities in the order of 5$\cdot 10^{11}$cm$^{-3}$ on the target's surface (not shown). 

\section{Summary}

Space and time resolved concentrations of Ar ($^3\mbox{P}_2$) metastable atoms in an atmospheric pressure radio-frequency micro-plasma jet were measured using tunable diode laser absorption spectroscopy. Metastable profiles as well as time resolved optical emission measurements reveal the Ar metastable dynamics and the discharge's filamentary structure. Spectral line profiles were recorded, allowing absolute metastable density calculations and reveal high electron densities inside the filaments. Additionally, Ar metastable densities in the order of 1$\cdot 10^{13}$cm$^{-3}$ have been estimated. The lifetime as well as the shape of the filaments was characterized by fast ICCD camera measurements. Highly time resolved TDLAS records show the expansion of the discharge after the ignition with a precision of 40\,ns. The spatial distribution of the filaments is characterized with and without a target in front of the jet, showing that the target potential and therewith the electric field distribution substantially changes the filaments' expansion. Together with the detailed analysis of the ignition phase and the discharge's behavior under pulsed operation, the results give an insight into the excitation mechanisms. The results show ways to enhance or reduce surface and volume reactive processes, which is interesting for a variety of applications like the deposition of coatings or medical purposes.

\ack
This project is supported by DFG (German Science Foundation) within the framework of the Research Unit FOR1123 (Projects A2, A4 \& C1) and the Research Department 'Plasmas with Complex Interactions' at Ruhr-University Bochum.

\newpage
\clearpage


\end{document}